\documentclass[letterpaper,10pt,conference]{ieeeconf}
\pdfoutput=1

\usepackage{color}
\usepackage{times}
\usepackage{flushend}
\usepackage[dvipsnames]{xcolor}
\usepackage{amsmath, mathtools}
\usepackage{amssymb}
\usepackage{graphicx}
\usepackage{epsfig}
\usepackage{bm}
\usepackage{url}
\usepackage[font=footnotesize, skip=4pt]{caption}
\usepackage{subfig}
\usepackage{amsthm}

\newtheorem{definition}{Definition}
\newtheorem{lemma}{Lemma}
\newtheorem{theorem}{Theorem}

\newtheorem{assumption}{Assumption}

\IEEEoverridecommandlockouts
\usepackage{tikz}
\usetikzlibrary{shapes,arrows}
\usetikzlibrary{bayesnet}
\usetikzlibrary{positioning}
\tikzstyle{status} = [rectangle, draw=black, text centered, anchor=north, text=black, minimum width=9em, minimum height=3em, node distance=6ex and 7em]
\tikzstyle{line} = [draw,thick,-latex]
\tikzstyle{transition} = [font=\small]

\makeatletter
\newcommand{\revisionhistory}[1]{%
\@ifundefined{showrevisionhistory}{\relax}{%
{#1}%
}%
}
\makeatother

\begin{document}


\title{\LARGE \bf How Peer Effects Influence Energy Consumption}
\author{Datong P. Zhou$^{\ast\dagger}$, Mardavij Roozbehani$^\dagger$, Munther A. Dahleh$^\dagger$, and Claire J. Tomlin$^\star$
\thanks{$^\ast$Department of Mechanical Engineering, University of California, Berkeley, USA. {\tt\footnotesize datong.zhou@berkeley.edu}}
\thanks{$^\dagger$Laboratory for Information and Decision Systems, MIT, Cambridge, USA.
{\tt\footnotesize [datong,mardavij,dahleh]@mit.edu}}
\thanks{$^\star$Department of Electrical Engineering and Computer Sciences, University of California, Berkeley, USA.
{\tt\footnotesize tomlin@eecs.berkeley.edu}}%
\thanks{This work has been supported in part by the National Science Foundation under CPS:FORCES (CNS-1239166) and CEC Grant 15-083.}
}

\maketitle
\thispagestyle{empty}
\pagestyle{empty}

\begin{abstract}
This paper analyzes the impact of peer effects on electricity consumption of a network of rational, utility-maximizing users. Users derive utility from consuming electricity as well as consuming less energy than their neighbors. However, a disutility is incurred for consuming more than their neighbors. To maximize the profit of the load-serving entity that provides electricity to such users, we develop a two-stage game-theoretic model, where the entity sets the prices in the first stage. In the second stage, consumers decide on their demand in response to the observed price set in the first stage so as to maximize their utility. To this end, we derive theoretical statements under which such peer effects reduce aggregate user consumption. Further, we obtain expressions for the resulting electricity consumption and profit of the load serving entity for the case of perfect price discrimination and a single price under complete information, and approximations under incomplete information. Simulations suggest that exposing only a selected subset of all users to peer effects maximizes the entity's profit.
\end{abstract}


%
%


\section{Introduction}
\label{sec:Introduction}
Energy efficiency programs have emerged as a viable resource to yield economic benefits to utility systems and to reduce the amount of greenhouse gas emissions. Demand-side management aims to modify consumer demand through financial incentive schemes and to induce behavioral changes through education. Specifically, users are offered rewards to conserve energy during peak hours or to shift usage to off-peak times. With communications and information technology constantly improving, which are characteristic elements of today's smart grid, demand-side management technologies are becoming increasingly feasible.

Previous academic work by psychologists, political scientists, and behavioral economists has found that social comparisons can have a significant impact on people's behavior, exploiting the willingness of individuals to conform to a standard, receive social acclaim, or simply the belief that other people's choices are informative in the presence of limited or imperfect information \cite{Akerlof:2000aa, Mani:2013aa}. 

Motivated by this line of academic work and the pressing need to improve energy efficiency, various companies and groups, for instance \texttt{OPOWER}, have conducted randomized control trials to investigate the impact of peer effects on energy consumption of residential households by sending out quarterly energy reports (so called \textit{Home Energy Reports}) to users with a comparison of their usage to their closest neighbors \cite{Ayres:2009aa}. While all experiments unanimously found an average reduction among the highest consuming users of around 1-2$\%$ \cite{Allcott:2011aa}, ambiguous results were found among low consumers, with one study reporting a ``boomerang effect'', that is, an increase of energy demand among the most efficient households.

Network effects in social networks and platforms often exhibit positive externalities, capturing the intuitive fact that an increased amount of platform activity promotes a local increase in platform activity. From a game-theoretic perspective, it is known that an analysis of games under such strategic complements admits well-behaved solutions if utility functions are supermodular with parameters drawn from a lattice \cite{Topkis:1998aa, Zhou:1994aa}. Examples for such games can be found in modeling technology adoption, human capital decisions, and criminal and social networks \cite{Calvo-Armengol:2005aa}. The opposite effect, that is, in games of strategic substitutes where an increased amount of activity leads to local reductions of activity, is observed in information sharing and the provision of public goods \cite{Bramoulle:2007aa}. However, since utility functions in this setting tend to lose the feature of supermodularity, finding equilibria is an inherently hard problem \cite{Jackson:2014aa}, and so these settings have been significantly less studied.

In an attempt to characterize the most influential players in a network, \cite{Ballester:2006aa} develops a quadratic model with continuous action spaces, a parameterization which we employ in this paper. Other research directions aiming at understanding the impact of network effects on social phenomena include diffusion models for the spread of information with the goal of influence maximization \cite{Kempe:2003aa}, repeated games to learn user interactions over time \cite{Acemoglu:2011aa}, or the analysis of systemic risk and stability \cite{Acemoglu:2015aa} in financial networks. The problem of profit maximization of a monopolist selling a divisible good, which is closely related with our work, has been investigated in \cite{Candogan:2012aa}, where the authors assume a constant marginal cost of production. However, to the best of our knowledge, a modeling approach for the impact of peer effects on energy consumption, whose generation typically has quadratic marginal cost, has yet to be formulated. 

In this paper, we propose a two-stage game-theoretic model for the energy consumption of a network of users, serviced by the load-serving entity that is obligated to cover the households' energy demand at all times. We analytically solve for the equilibria of this game under full information of the network structure and users' parameters to characterize the influence of peer effects on aggregate consumption and utility profit, for both the case of perfect price discrimination and a single price valid for all users. For the case of incomplete information, we obtain approximations of the utility's profit, user consumptions, and the optimal pricing scheme. Further, we analyze the profit-maximization problem by selecting the best subset of users to be exposed to peer effects, and present a heuristic solution to this NP-hard selection problem. Lastly, we provide theoretical statements on the properties of users which ensure that the consumption under peer effects is reduced.

The remainder of this paper is organized as follows: Section \ref{sec:Game_Theoretical_Model} presents the two-stage game-theoretic model between the utility and the network of consumers and derives consumption and price equilibria. Based on this model, Section \ref{sec:Theoretical_Statements} presents various theorems on the reduction of consumption in response to the peer effect as well as on the effect of uncertainty of the network structure on the optimal profit. Section \ref{sec:comparison_pricing_schemes} compares the utility's profit under the pricing schemes derived in Section \ref{sec:Game_Theoretical_Model}. Next, the challenge of maximizing the utility's profit by imposing a binary constraint on the number of users exposed to peer effects is formulated and solved with a heuristic approach in Section \ref{sec:profit_max_fixed_price}. Section \ref{sec:Conclusion} concludes the paper. All proofs are relegated to the Appendix.

%
%


\section{Game-Theoretic Model}
\label{sec:Game_Theoretical_Model}
\subsection{Players}
Define the set of consumers as $\mathcal{I}=\lbrace 1, \ldots, n\rbrace$. Let $W \in \mathbb{R}^{n\times n}$ define the interaction matrix which describes the network links and strengths between users. More precisely, let $w_{ij} \in [0, 1]$ denote the strength of influence of user $j$ on $i$. We assume $w_{ii}=0~\forall~i\in\mathcal{I}$ and normalize the row sums, $\sum_{j\in\mathcal{I}}w_{ij}=1~\forall~i\in\mathcal{I}$. Each element $w_{ij}>0$ in $W$ corresponds to a directed edge from agent $j$ to agent $i$, that is, the adjacency matrix $G$ of the resulting directed graph is the transpose of $W$. Each user $i$ derives a utility $u_i\in\mathbb{R}$ from consuming $x_i$ units of electricity as follows:
\begin{align}\label{eq:user_utility_function_peer_effect}
u_i &= a_i x_i - b_i x_i^2 - p_i x_i + \gamma_i x_i \left(\sum_{j\in\mathcal{I}}w_{ij} x_j - x_i \right).
\end{align}
In \eqref{eq:user_utility_function_peer_effect}, $a_i$ and $b_i$ denote user-specific parameters to describe the concave and increasing direct utility from consuming $x_i$ units of electricity, and $p_i$ denotes the unit price set by the utility. The last term captures the strategic complementarity between user $i$ and its neighbors. It is positive if user $i$ consumes less than the average of its neighbors, and vice versa. The difference between the average consumption and the user consumption is scaled by a proportionality constant $\gamma_i$ and the consumption level $x_i$.

Since each user consumes $x_i$ units of electricity at unit price $p_i$, the utility's profit reads as follows:
\begin{align}\label{eq:utility_profit_function}
\Pi = \sum_{i\in\mathcal{I}} p_i x_i - c_i x_i^2,
\end{align}
where the marginal cost of production $2c_i x_i$ is assumed to be linear in the production quantity $x_i$, which is a standard and often made assumption. For expositional ease, we further assume that the utility generates electricity itself and does not procure it from the wholesale electricity market. Relaxing this assumption would introduce uncertainty in wholesale prices, a problem which is outside the scope of this paper.

\subsection{Two-Stage Game}
To model the hierarchy between the utility, which acts as a monopolist that has the power to set prices, and the users, we formulate a two-stage game as follows:
\begin{enumerate}
\item The utility determines the optimal price $\mathbf{p}^\ast$ so as to maximize its profit by taking into account users' consumption decisions as a function of any particular price vector $\mathbf{p}$, that is,
\begin{align}
\mathbf{p}^\ast = \arg\max_{\mathbf{p}\geq\mathbf{0}}~\sum_{i\in\mathcal{I}} p_i x_i(p_i) - c_i x_i^2(p_i)
\end{align}
\item Each agent observes the price $p_i^\ast$ and $\mathbf{x}_{-i}$ and consumes $x_i^\ast$ units of electricity so as to maximize her utility, that is, $x_i^\ast = \arg\max_{x_i\geq 0}u_i(x_i, \mathbf{x}_{-i}, \gamma_i, W)$.
\end{enumerate}

We will solve this two-stage game by finding a subgame perfect equilibrium for the cases of perfect price discrimination and a single price for all users. We also differentiate between the full-information case where the utility has knowledge about all $\lbrace a_i\rbrace_{i=1}^n$ and $\lbrace b_i\rbrace_{i=1}^n$, and the case in which only their expectations $\mathbb{E}[a]$ and $\mathbb{E}[b]$ are known.

\subsection{Subgame-Perfect Equilibrium}

\begin{assumption}\label{as:a_strictly_greater_than_p}
$a_i > p_i~\text{and}~b_i > \gamma_i~\forall~i\in\mathcal{I}$.
\end{assumption}

\begin{theorem}\label{thm:second_stage_equilibrium}
Given the price vector $\mathbf{p}$ and consumption vector $\mathbf{x}_{-i}$, the utility maximizing response of user $i$ is
\begin{equation}\label{eq:consumption_equilibrium}
x_i^\ast = \frac{a_i-p_i + \gamma_i\sum_{j\in\mathcal{I}} w_{ij}x_j}{2(b_i+\gamma_i)}.
\end{equation}
Further, $\lbrace x_1^\ast, \ldots, x_n^\ast \rbrace$ constitute a unique Nash Equilibrium of the second stage game.
\end{theorem}
Recall that $w_{ii}=0~\forall~i\in\mathcal{I}$, which allows the right hand side of \eqref{eq:consumption_equilibrium} to depend on $\mathbf{x}_{-i}$ only. Assumption \ref{as:a_strictly_greater_than_p} is necessary to ensure that \eqref{eq:consumption_equilibrium} is indeed a maximum attained at a non-negative value. With the definitions $B:=\text{diag}\left(2b_1, \ldots, 2b_n\right)$ and $\Gamma:=\text{diag}\left(\gamma_1, \ldots, \gamma_n\right)$, \eqref{eq:consumption_equilibrium} can be rewritten as
\begin{equation}\label{eq:consumption_equilibrium_rewritten}
\mathbf{x}^\ast = \left(B + 2\Gamma - \Gamma W\right)^{-1} (\mathbf{a}-\mathbf{p}).
\end{equation}

\begin{definition}[Katz-Bonacich Centrality \cite{Katz:1953aa, Bonacich:1987aa}]\label{def:katz_centrality}
Given the adjacency matrix $G$, the weight vector $\mathbf{w}$, and the scalar $0\leq\alpha<1/\rho(G)$, where $\rho(G)$ denotes the spectral radius of $G$, the weighted Katz-Bonacich Centrality is defined as 
\begin{align}\label{eq:katz_centrality}
\mathcal{K}_{\mathbf{w}}(G, \alpha) = \left(I - \alpha G\right)^{-1} \mathbf{w} = \sum_{k=0}^\infty (\alpha G)^k \mathbf{w}.
\end{align}
The centrality of a particular node $i$ can be interpreted as the sum of total number of walks from $i$ to its neighbors discounted exponentially by $\alpha$ and weighted by $w_i$.
\end{definition}

For the special case $\gamma_1 = \ldots = \gamma_n = \gamma$, and noting that $G = W^\top$, \eqref{eq:consumption_equilibrium_rewritten} can be rewritten in terms of the weighted Katz-Bonacich Centrality:
\begin{align*}
\mathbf{x}^\ast &= (B + 2\gamma I)^{-1}\left( I - \gamma W^\top(B + 2\gamma I)^{-1} \right)^{-1}(\mathbf{a}-\mathbf{p})\\
&= (B + 2\gamma I)^{-1}\mathcal{K}_{\mathbf{a}-\mathbf{p}}(W^\top(B + 2\gamma I)^{-1}, \gamma)
\end{align*}
We note that  $\left(B + 2\Gamma - \Gamma W\right)$ is strictly diagonally dominant for all $\gamma \geq 0$, with positive diagonal entries. The Gershgorin Circle Theorem then states that all its eigenvalues are strictly positive, from which invertibility follows.

We first focus on the full information case and present the equilibria in Theorems \ref{thm:opt_price_under_perfect_price_discrimination} and \ref{thm:profit_maximizing_uniform_price}. Let $C = \text{diag}(c_1, \ldots, c_n)$.

\begin{theorem}\label{thm:opt_price_under_perfect_price_discrimination}
Under perfect price discrimination, the profit-maximizing solution $\mathbf{p}^\ast$ to the first stage game is
\begin{align}\label{eq:optimal_price_under_ppd}
\mathbf{p}^\ast &= \underbrace{\frac{\mathbf{a}}{2}}_{(1)} + \underbrace{CZ\frac{\mathbf{a}}{2}}_{(2)} - \underbrace{W^\top \Gamma Z\frac{\mathbf{a}}{4}}_{(3)} +\underbrace{\Gamma WZ\frac{\mathbf{a}}{4}}_{(4)},\\
Z &= \left[2\Gamma + B + C - \left(\frac{W^\top \Gamma}{2} + \frac{\Gamma W}{2}\right) \right]^{-1}.\nonumber
\end{align}
The four components are interpreted as follows:
\begin{enumerate}
\item A constant term $a_i/2$, c.f. $a_i$ in \eqref{eq:user_utility_function_peer_effect},
\item An additional cost that correlates with cost $c_i$,
\item An incentive for strongly influential users $W^\top \Gamma$,
\item An additional cost for strongly influenced users $\Gamma W$.
\end{enumerate}
The optimal consumption under this policy is
\begin{align}\label{eq:opt_consumption_no_price}
\mathbf{x}^\ast = \left(C + B + 2\Gamma - \frac{W^\top\Gamma}{2}-\frac{\Gamma W}{2}\right)^{-1}\frac{\mathbf{a}}{2}.
\end{align}
For the special case of symmetric networks, i.e. $W=W^\top$, the optimal profit $\Pi^\ast$ becomes
\begin{align}
\Pi^\ast = \frac{1}{4}\mathbf{a}^\top (C + B + 2\Gamma - \Gamma W)^{-1} \mathbf{a}.
\end{align}
\end{theorem}

\begin{theorem}\label{thm:profit_maximizing_uniform_price}
Under complete information, i.e. the utility knows $a_i$ and $b_i~\forall~i\in\mathcal{I}$, the profit-maximizing single price $p_u^\ast$ is
\begin{align}\label{eq:unif_price_complete}
p_u^\ast = \left[1 - \frac{\mathbf{1}^\top A^{-1}\mathbf{1}}{2\cdot\mathbf{1}^\top\left( A^{-1} + A^{-1}CA^{-1} \right)\mathbf{1}}\right]\bar{a}
\end{align}
and the consumption equilibrium writes
\begin{align}\label{eq:unif_consumption_complete}
\mathbf{x}^\ast = A^{-1}\left[\mathbf{a} - \left( 1 - \frac{\mathbf{1}^\top A^{-1}\mathbf{1}}{2\cdot\mathbf{1}^\top\left( A^{-1} + A^{-1}CA^{-1} \right)\mathbf{1}} \right)\bar{a}\mathbf{1} \right],
\end{align}
where $A = B + 2\Gamma - \Gamma W$ and $\bar{a} = \sum_{i=1}^n a_i/n$.
\end{theorem}

\begin{lemma}\label{lem:uniform_price_symmetric_properties}
For symmetric networks, i.e. $W=W^\top$, the single profit-maximizing price \eqref{eq:unif_price_complete} and its corresponding consumption \eqref{eq:unif_consumption_complete} simplify to
\begin{subequations}
\begin{align}
p_u^\ast &= \frac{1}{n}\sum_{i=1}^n p_i^\ast \label{eq:unif_price_symmetric}\\
\mathbf{x}_u^\ast &= \left(B + 2\Gamma -\Gamma W\right)^{-1}(\mathbf{a}-\bar{a}\mathbf{1}) \label{eq:unif_cons_symmetric}\\
&\quad+ \left(C + B + 2\Gamma - \Gamma W\right)^{-1}\frac{\bar{a}}{2}\mathbf{1} \nonumber.
\end{align}
\end{subequations}
\end{lemma}

By construction of the optimal prices and consumptions, the optimal profit under a single price is less than under perfect price discrimination, that is, $\Pi_u^\ast \geq \Pi^\ast$.

Next, for the incomplete information scenario and additional assumptions $W=W^\top$ and $C=cI$, the utility can approximate the profit-maximizing price as in Theorem \ref{thm:opt_price_incomplete_information}.

\begin{theorem}\label{thm:opt_price_incomplete_information}
In the case of incomplete information, that is, only the expectations of $\lbrace a_i\rbrace_{i=1}^n$ and $\lbrace b_i\rbrace_{i=1}^n$ are known and denoted with $\mathbb{E}[a]$ and $\mathbb{E}[b]$, the optimal single profit-maximizing price $\tilde{p}_u^\ast$ and the expected corresponding consumption equilibrium $\mathbb{E}[\tilde{x}_i]$ are bounded below by
\begin{subequations}
\begin{align}
\tilde{p}_u^\ast &\geq \frac{\mathbb{E}[a]}{2}\left[ 1 + \frac{c}{n}\mathbf{1}^\top \left[ 2\Gamma + (2\mathbb{E}[b]+c)I - \Gamma W \right]^{-1}\mathbf{1}\right], \label{eq:unif_price_incomplete}\\
\mathbb{E}[\tilde{x}_i] &\geq \frac{\mathbb{E}[a]-\tilde{p}_{u,\text{LB}}^\ast}{n}\cdot\mathbf{1}^\top \left( 2\Gamma + 2\mathbb{E}[b]I -\Gamma W \right)^{-1} \mathbf{1.}\label{eq:unif_consumption_incomplete}
\end{align}
\end{subequations}
where $\tilde{p}_{u,\text{LB}}^\ast$ denotes the lower bound on the single profit-maximizing price $\tilde{p}_u^\ast$ \eqref{eq:unif_price_incomplete}.
\end{theorem}

\begin{theorem}[Profit Maximizing Price without Peer Effects]\label{thm:opt_price_uniform_price}
In the case of incomplete information and in the absence of any peer effects, the single profit-maximizing price $\hat{p}^\ast$ and the expected user consumption $\mathbb{E}[\hat{x}_i]$ are
\begin{subequations}
\begin{align}
\hat{p}^\ast &= \frac{\mathbb{E}[b]+c}{2\mathbb{E}[b]+c}\mathbb{E}[a], \label{eq:optimal_uniform_price}\\
\mathbb{E}[\hat{x}_i^\ast] &= \frac{\mathbb{E}[a]}{2(2\mathbb{E}[b]+c)}\quad~\forall~i\in\mathcal{I}. \label{eq:optimal_expected_consumption_unif_price}
\end{align}
\end{subequations}
\end{theorem}



%
%


\section{Theoretical Statements}
\label{sec:Theoretical_Statements}
We next seek to analyze under what conditions the aggregate consumption across all users is less than in the absence of peer effects, which is a desirable goal from the energy efficiency perspective.
\begin{theorem}\label{thm:optimal_consumption_decreasing_gamma}
If $a_i =: a$, $b_i =: b$, and $\gamma_i =: \gamma~\forall~i\in\mathcal{I}$, and Assumption \ref{as:a_strictly_greater_than_p} holds, then $x_i^\ast$ \eqref{eq:consumption_equilibrium} is strictly monotonically decreasing in $\gamma$, independent of the network topology $W$. 
\end{theorem}
Theorem \ref{thm:optimal_consumption_decreasing_gamma} is interesting because identical consumers will reduce their optimal consumption compared to the case of no peer effects, even though $x^\ast_i=x^\ast_j~\forall~i,j\in\mathcal{I}$ and hence the peer effect term $\gamma_i x_i \left(\sum_{j\in\mathcal{I}}w_{ij} x_j - x_i \right)$ is zero.

\begin{theorem}[\textbf{Influence of High Consumer}]\label{thm:high_consumer_influence}
Given that $w_{ij}=\left( \sum_{j\in\mathcal{I}}1_{w_{ij}>0} \right)^{-1}~\forall~i\in\mathcal{I}$, that is, all connections are of equal weight, and $b_i =: b$ and $\gamma_i =: \gamma~\forall~i\in\mathcal{I}$. Define the set of users $\mathcal{N} := \lbrace i\in\mathcal{I}\setminus j \rbrace$ with the characteristic $a_i-p_i =: \alpha~\forall~i\in\mathcal{N}$. Further, let $j$ be a ``high consumer'', that is, $a_j-p_j =:\bar{\alpha} > n\alpha$. Denote the set of all neighbors of $j$ as $\mathcal{C}_j:=\lbrace i\in\mathcal{N}~|~w_{ij}> 0 \rbrace$. Then, independent of the network topology, for all users $i \in \mathcal{C}_j$, $x_i^\ast$ is increasing for small enough values of $\gamma$ whereas $x_j^\ast$ is strictly monotonically decreasing in $\gamma$.
\end{theorem}
Let $m_i$ denote the number of neighbors of consumer $i$. Theorem \ref{thm:high_consumer_influence} can be restated as in Lemma \ref{lem:restate_high_consumer_influence}.
\begin{lemma}\label{lem:restate_high_consumer_influence}
$x_i^\ast,~i\in\mathcal{C}_j$ is increasing for small enough values of $\gamma$ if $\bar{\alpha} \geq m_j+1$. Equivalently, if $\bar{\alpha} = k\alpha,~k\in\mathbb{N}$, only the subset $\lbrace i\in\mathcal{C}_j~|~m_i \leq k-1\rbrace$, i.e. the set of users with fewer than $k-1$ neighbors, shows an initial increase in consumption as a function of $\gamma$.
\end{lemma}
Theorem \ref{thm:high_consumer_influence} and Lemma \ref{lem:restate_high_consumer_influence} describe conditions on the average consumption of any particular user's neighbors to observe a ``boomerang effect'', given there is a unique ``high'' consumer among a pool of users of identical characteristics.

\begin{theorem}[\textbf{Targeted Peer Effects}]\label{thm:targeted_peer_effects}
For a general setting of $n\geq 2$ users with non-identical parameters $a_i, b_i$ and a fixed price $p$ among all users, exposing exactly two connected users to the peer effect, w.l.o.g. referred to as users ``1'' and ``2'', reduces the sum of their consumptions under the following conditions:
\begin{subequations}
\begin{align}\label{eq:sum_of_consumptions_conditions}
b_1 &\leq \frac{(a_1 - p)\left[ 4(b_2+\gamma) - \gamma w_{12}w_{21} \right]}{4(b_2+\gamma)\sum\limits_{j=3}^n w_{1j}x_j + 2w_{12}\bigg( a_2 - p + \gamma\sum\limits_{j=3}^n w_{2j}x_j \bigg)} \\
b_2 &\leq \frac{(a_2 - p)\left[ 4(b_1+\gamma) - \gamma w_{12}w_{21} \right]}{4(b_1+\gamma)\sum\limits_{j=3}^n w_{2j}x_j + 2w_{21}\bigg( a_1 - p + \gamma\sum\limits_{j=3}^n w_{1j}x_j \bigg)}
\end{align}
\end{subequations}
where $x_j,~j\in\lbrace 3, \ldots, n\rbrace$ is given by $x_j = (a_j-p)/(2b_j)$.
For the special case of $n=2$, this condition reads
\begin{align*}
b_1 \leq \frac{(a_1-p)\left( 4b_2+3\gamma \right)}{2(a_2-p)}\quad\text{and}\quad b_2 \leq \frac{(a_2-p)\left( 4b_1+3\gamma \right)}{2(a_1-p)} 
\end{align*}
\end{theorem}
Theorem \ref{thm:targeted_peer_effects} states that if two connected users both receive notifications of their neighbors' consumption, the sum of their consumptions decreases as long as they are not ``too different'' from each other and their neighbors. Thus, the total consumption of a network of users correlates negatively with the number of users given the treatment. Analogous bounds can be found for exposing more than two users to the peer effect at the expense of notational ease.

Finally, we investigate the case of incomplete information about the network structure for the case of symmetric networks, i.e. $W=W^\top$. It is assumed that the monopolist only knows an approximation of $W$, denoted with $\tilde{W}$, where $\tilde{W}=\tilde{W}^\top$. Under perfect price discrimination, the utility can set profit-maximizing prices in the first stage of the game, assuming that users' consumption $\mathbf{\tilde{x}}$ in the first stage is determined according to $\tilde{W}$. The real consumption $\mathbf{x}^\ast$, however, follows the actual $W$ (which is unknown to the utility). Theorem \ref{thm:uncertainty_network_structure} provides a lower bound on the ratio of the optimal expected profit under network uncertainty to the profit obtainable under perfect network information.
\begin{theorem}[\textbf{Uncertainty in $\boldsymbol{W}$}]\label{thm:uncertainty_network_structure}
Assume that $W=W^\top$ and $\Gamma = \gamma I, \gamma \geq 0$. If the monopolist has access only to the estimate $\tilde{W}$ with $\tilde{W}=\tilde{W}^\top$, then, under perfect price discrimination, the ratio of optimal expected profit $\tilde{\Pi}^\ast$ to profit $\Pi^\ast$ under perfect knowledge of $W$ is bounded below:
\begin{align}\label{eq:bound_profit_uncertainty}
\frac{\tilde{\Pi}^\ast}{\Pi^\ast} \geq \frac{\lambda_{\mathrm{min}}(C+B + 2\Gamma - \Gamma W)}{\lambda_{\mathrm{max}}(C+B + 2\Gamma - \Gamma W) + \gamma\Vert W-\tilde{W} \Vert_2},
\end{align}
where $\Vert\cdot\Vert_2$ is the Euclidian matrix norm.
\end{theorem}
For the edge case $\tilde{W} = 0$, we have $\Vert W\Vert_2 = 1$ due to the well-known fact that the maximal eigenvalue of an adjacency matrix is the degree of the graph. Due to row normalizations of $W$, the degree is $1$, which corresponds to the eigenvector $\mathbf{1}$ associated with eigenvalue 1. To qualitatively show that the bound \eqref{eq:bound_profit_uncertainty} becomes tighter as $\tilde{W}$ approaches $W$, observe that $\Vert W-\tilde{W}\Vert_2$ corresponds to the largest singular value of $W-\tilde{W}$, which is identical to its spectral radius because $W-\tilde{W}$ is Hermitian. Finally, the Gershgorin Circle Theorem states that every eigenvalue of $W-\tilde{W}$ lies within at least one of the disks that is centered at the origin, each of which has radius $R_i = \sum_{j\neq i}|w_{ij}-\tilde{w}_{ij}|$. As $w_{ij} \rightarrow \tilde{w}_{ij}$, $R_i \rightarrow 0$.

\begin{figure}[hbtp]
\vspace*{-0.3cm}
\centering
\includegraphics[scale=0.288]{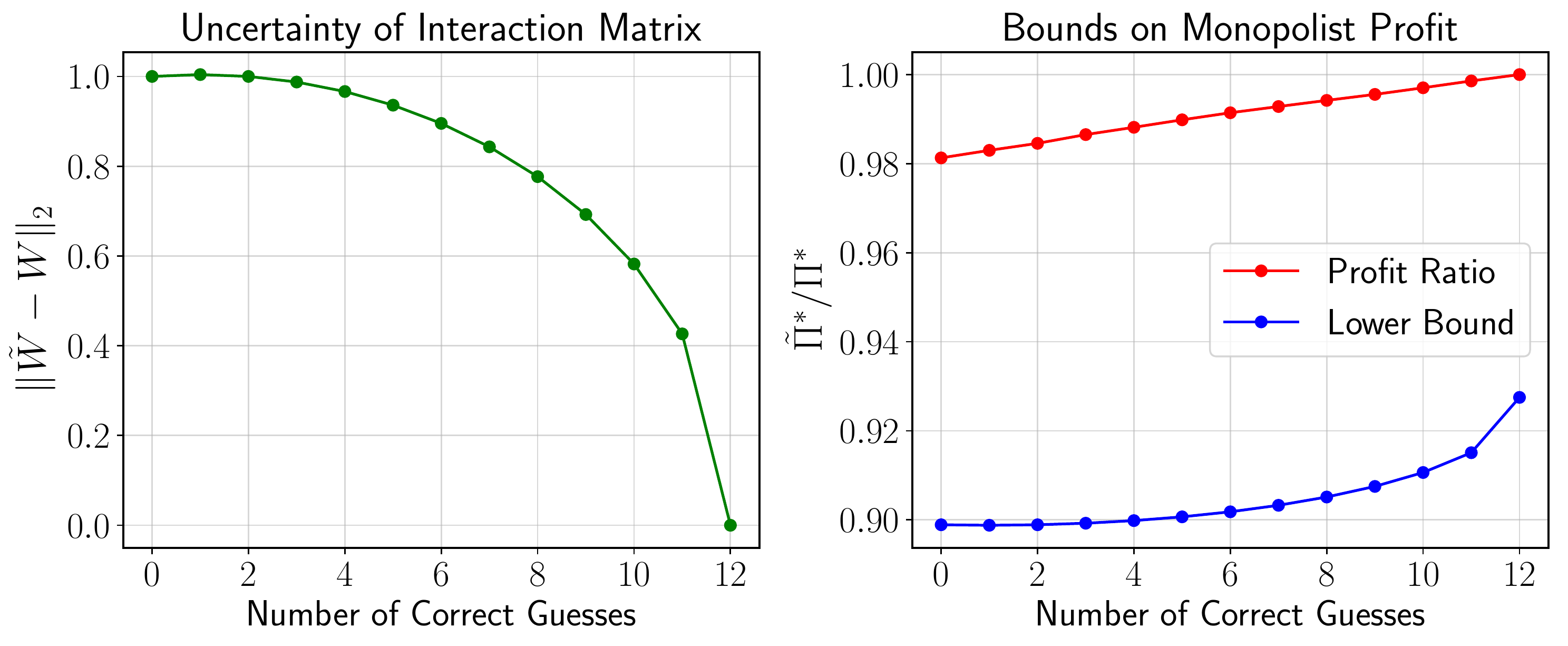}
\caption{$\Vert W-\tilde{W}\Vert_2$ for 12 fully connected users embedded in a network of $n=24$ customers, $\gamma=0.05$.}
\vspace*{-0.3cm}
\label{fig:interaction_matrix_norm}
\end{figure}

To illustrate the bound \eqref{eq:bound_profit_uncertainty}, let $n=24$ and $W\in\mathbb{R}^{24\times 24}$ be the ground truth interaction matrix of 12 randomly chosen, fully connected users, whose parameters $a_i, b_i,$ and $c_i~\forall~i\in\mathcal{I}$ are randomly drawn from appropriate distributions. Assuming that the monopolist knows that 12 out of 24 users are fully connected, we iterate through all $24\choose 12$ combinations and calculate $\Vert \tilde{W} - W\Vert_2$ and the profit bound \eqref{eq:bound_profit_uncertainty} as a function of the number of correct user assignments, where we take the mean across any particular number of correct assignments. As the number of correct assignments increases, the metric for the mismatch between $W$ and $\tilde{W}$, namely $\Vert \tilde{W} - W\Vert_2$ decreases, whereas the profit bound increases, see Figure \ref{fig:interaction_matrix_norm}.

\begin{theorem}[Efficiency]\label{thm:efficiency_consumption_equilibrium}
The consumption equilibrium $\mathbf{x}^\ast$ \eqref{eq:opt_consumption_no_price} is inefficient as the social welfare $\mathcal{S}$ attained at \eqref{eq:opt_consumption_no_price} is suboptimal. Specifically, $x_i^\ast < x_i^o~\forall~i\in\mathcal{I}$, where $\mathbf{x}^o$ denotes the consumption that maximizes social welfare, which reads
\begin{align}\label{eq:socially_optimal_consumption}
\mathbf{x}^o = \left( C + \frac{B}{2} + \Gamma - \frac{W^\top\Gamma}{2} - \frac{\Gamma W}{2} \right)^{-1}\frac{\mathbf{a}}{2}.
\end{align}
Allocating users per-unit subsidies $s_i = (b_i+\gamma_i)x_i^2/2$ (Pigouvian Subsidy) can restore the social optimum.
\end{theorem}




%
%

\section{Comparison of Pricing Schemes}\label{sec:comparison_pricing_schemes}
\subsection{Network Topologies}
\begin{figure}[hbtp]
\vspace{-0.25cm}
\begin{tikzpicture}
\node[latent]                                (f1) {1};
\node[latent, below=of f1]                   (f3) {3};
\node[latent, right=of f3]                   (f4) {4};
\node[latent, right=of f1]                   (f2) {2};

\node[latent, right=of f2, xshift=-0.3cm]    (r1) {1};
\node[latent, below=of r1]                   (r3) {3};
\node[latent, right=of r3]                   (r4) {4};
\node[latent, right=of r1]                   (r2) {2};

\node[latent, right=of r2, xshift=-0.3cm]    (s1) {1};
\node[latent, below=of s1]                   (s4) {3};
\node[latent, right=of s4]                   (s3) {4};
\node[latent, right=of s1]                   (s2) {2};

\path (f1) edge (f2);
\path (f1) edge (f3);
\path (f1) edge (f4);
\path (f2) edge (f3);
\path (f2) edge (f4);
\path (f3) edge (f4);

\path (r1) edge (r2);
\path (r1) edge (r3);
\path (r1) edge (r4);

\path (s1) edge (s2);
\path (s2) edge (s3);
\path (s3) edge (s4);
\path (s4) edge (s1);
\end{tikzpicture}
\vspace{0.0cm}
\caption{Basic network architectures for $n=4$: Fully connected, star, ring}
\label{fig:basic_networks}
\end{figure}
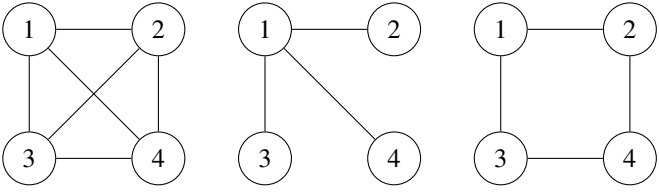

In the remainder of this paper, we assume users to be connected to each other through one of the basic network topologies displayed in Figure \ref{fig:basic_networks}.

\subsection{Simulation}
We now simulate the consumption and price equilibria as well as the profit of the monopolist as a function of the network strength parameter $\gamma$ under the following three pricing scenarios:
\begin{itemize}
\item \textbf{Case 1}: Monopolist has complete information of $\mathbf{a}$ and $\mathbf{b}$ and sets prices with perfect price discrimination \eqref{eq:optimal_price_under_ppd};
\item \textbf{Case 2}: Monopolist has complete information of $\mathbf{a}$ and $\mathbf{b}$ and sets the profit-maximizing single price \eqref{eq:unif_price_complete};
\item \textbf{Case 3}: Monopolist has access only to $\mathbb{E}[a]$ and $\mathbb{E}[b]$ and sets the lower bound on the single price \eqref{eq:unif_price_incomplete}.
\end{itemize}

\begin{figure}[hbtp]
\centering
\vspace*{-0.35cm}
\includegraphics[scale=0.345]{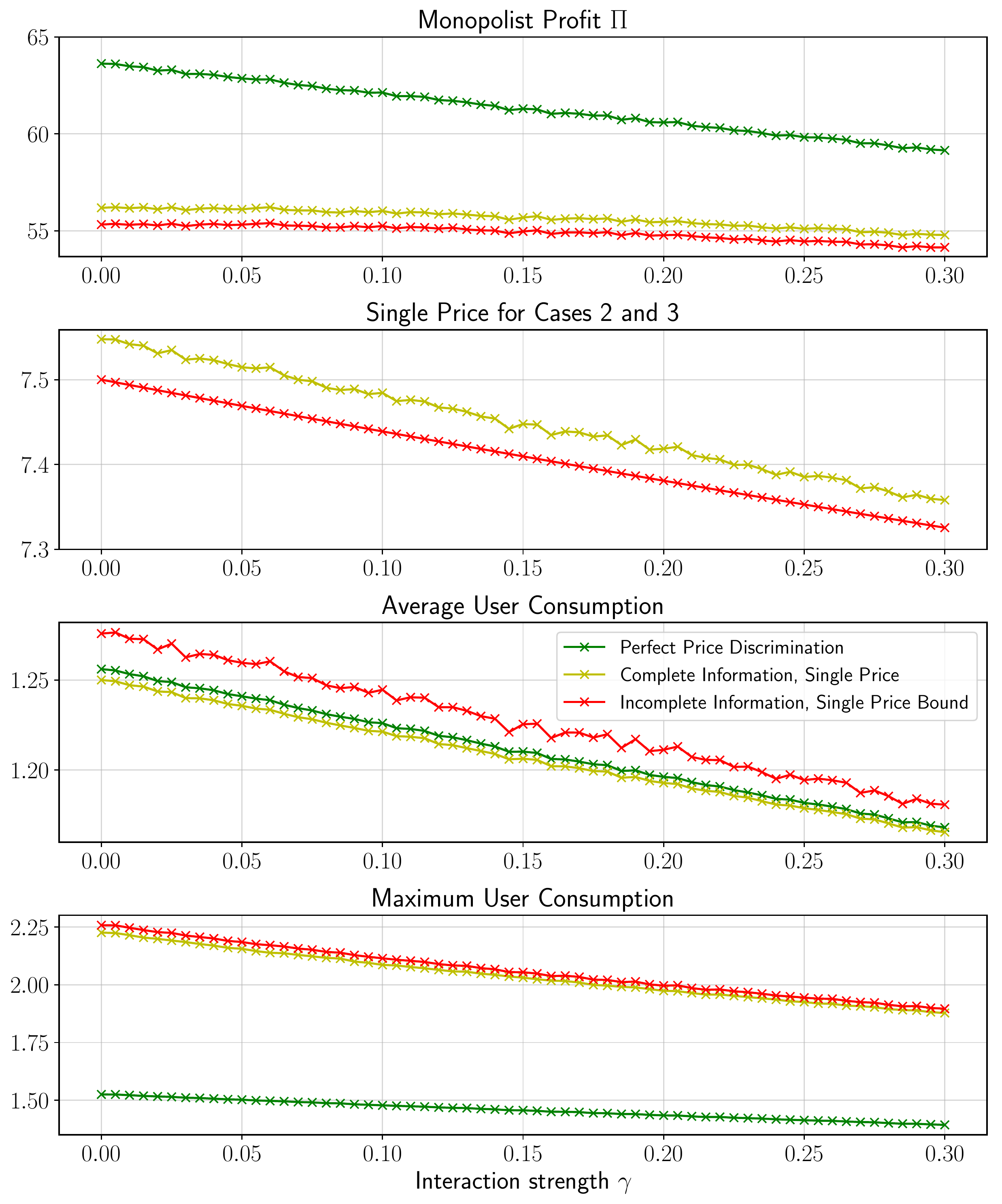}
\caption{Profit of monopolist, single prices \eqref{eq:unif_price_complete} and \eqref{eq:unif_price_incomplete}, average user consumption, and maximum consumption under perfect price discrimination (green), single pricing under complete information \eqref{eq:unif_price_complete} (yellow), and single pricing under incomplete information \eqref{eq:unif_price_incomplete} (red). 10,000 iterations, $a\sim\text{unif}[8,12],~b\sim\text{unif}[0.75, 1.25],~c_i=2~\forall~i\in\mathcal{I}$.}
\vspace*{-0.15cm}
\label{fig:Utility_Profit_Pricing_Comparison}
\end{figure}

We simulate a network of $n=10$ fully connected users with $a_i$ and $b_i$ randomly drawn from uniform distributions with support $[8, 12]$ and $[0.75, 1.25]$, respectively. The cost is set to $c_i=2$ for all users. As the results for the star and ring network are qualitatively similar to the fully connected network, we omit discussions of these cases. The optimal prices for each of the cases (1)-(3) are then calculated, which fixes the users' consumptions and the monopolist's profit. Repeating this process 10,000 times and taking the mean across all iterations yields the characteristics in Figure \ref{fig:Utility_Profit_Pricing_Comparison}.

As expected, the profit under perfect price discrimination \eqref{eq:optimal_price_under_ppd} exceeds the profit obtained with cases (2) and (3), where, somewhat surprisingly, setting the lower bound on the prices (case (3)) does not give up too much profit, compared to case (2). This indicates that the lower bound on the optimal price \eqref{eq:unif_price_incomplete} is ``close'' to the actual optimum, which is proven by the second subplot, from which it follows that \eqref{eq:unif_price_incomplete} falls short of \eqref{eq:unif_price_complete} by less than $<1\%$.

Consequently, the lower price bound \eqref{eq:unif_price_incomplete} results in a higher average user consumption than in case (2), which directly follows from the consumption equilibrium \eqref{eq:consumption_equilibrium}. The average user consumption under perfect price discrimination is sandwiched between cases (2) and (3).

Lastly, the maximum user consumption for perfect price discrimination is about $30\%$ lower than in cases (2) and (3), which has beneficial side-effects on grid operation. This observation also motivates the heuristic user-selection algorithm presented in the next section.

\section{Profit Maximization with User Selection}
\label{sec:profit_max_fixed_price}

\subsection{Problem Formulation}
We now seek to answer the following question: Given the single, exogenous price $p$ and the parameters $\lbrace a_i\rbrace_{i=1}^n$ and $\lbrace b_i\rbrace_{i=1}^n$ sampled from distributions with means $\mathbb{E}[a]$ and $\mathbb{E}[b]$, respectively and are known to the monopolist, which users should be targeted to maximize profit? This situation can arise if the utility is obligated to charge customers at a rate $p$ per unit of electricity and only wants to spend a limited budget on informing users about their peers' behavior. In other words, which best subset of all users should be exposed to the peer effect such that the utility achieves maximum profit under exogenous price $p$? The profit maximizing problem of the utility thus writes
\begin{equation}\label{eq:utility_targeting_optimization_problem}
\begin{aligned}
& \underset{\delta_1, \ldots, \delta_n}{\text{maximize}}
& & \sum_{i=1}^n p x_i - c_i x_i^2 \\
& \text{subject to}
& & \mathbf{x} = \left(B + 2\Delta \Gamma - \Delta \Gamma W\right)^{-1}(\mathbf{a}-p\mathbf{1}) \\
& & & \sum_{i=1}^n \delta_i = m,\quad \delta_i\in\lbrace 0, 1\rbrace \\
\end{aligned}
\end{equation}
where $\Delta = \text{diag}(\delta_1, \ldots, \delta_n)$ and $\delta_i=1$ and $\delta_i = 0$ denote that user $i$ is targeted or non-targeted, respectively. This is an NP-hard Mixed Integer Quadratically Constrained Program (MIQCP) due to the binary constraint to expose exactly $m$, $0\leq m \leq n$ users to the network effect and the quadratic objective, and so \eqref{eq:utility_targeting_optimization_problem} does not admit a closed form solution. An analytical solution requires exhaustive search, which is computationally infeasible for any real network of users. Therefore, we resort to the following heuristic which was hinted at at the end of Section \ref{sec:comparison_pricing_schemes}: Given the user parameters $\mathbf{a}$ and $\mathbf{b}$ and the single price $p$, we first compute the consumptions in the absence of any network effects, denoted with $\tilde{\mathbf{x}} = B^{-1}(\mathbf{a}-p\mathbf{1})$. Next, we calculate the optimal consumptions with the expectations of $\mathbb{E}[a]$ and $\mathbb{E}[b]$, which we denote with $\mathbb{E}[x]$. Lastly, the pairwise differences $|\mathbb{E}[x] - \tilde{x}_i|$ are put into a sorted list, and the heuristic selection algorithm returns the indices of the $m$ largest values in this list. That is, $\Delta_h = \text{diag}(\delta_{h,1}, \ldots, \delta_{h,n})$, where $\delta_{h,i} = 1$ if consumer $i$ belongs to the set of the $m$ largest $|\mathbb{E}[x]-\tilde{x}_i|$, and $\delta_{h,i}=0$ otherwise.

The idea of this heuristic is motivated by Theorem \ref{thm:high_consumer_influence}, according to which a high consumer in a network of low consumers can result in a consumption increase of low consumers. Since the user parameters are sampled from a finite distribution, a single price on non-identical users always results in suboptimal profit, but approaches optimality as users become more similar. Exposing the highest and lowest consumers (measured against $\mathbb{E}[x]$) to the network effect nudges high users (low users) to consume less (more), thereby making the users more similar in their consumption, which in turn increases the utility's profit.

Further, the fact that the maximum user consumption under perfect price discrimination (which achieves notably better profit than single pricing, see Figure \ref{fig:Utility_Profit_Pricing_Comparison}) is about 30~\% lower than under single pricing corroborates the notion of exposing high consumers to the peer effects. According to Theorem \ref{thm:high_consumer_influence}, such users reduce their consumption in response to the peer effect, which reduces the maximum user consumption to increase profit.

The utility needs to find the sweet spot between the following two extremes: Targeting too few users results in a suboptimal increase in profit. On the other hand, according to Theorem \ref{thm:targeted_peer_effects}, targeting too many users leads to an overall consumption decrease because targeting a customer whose neighbors are already exposed to the network causes the neighbors to reduce their consumption further.

Note that this heuristic neither takes into account the interaction matrix $W$ nor the fact whether the deviation of the actual consumption from the expected one is positive or negative, and so it could be improved by running a classification algorithm on the features $|\mathbb{E}[x]-\tilde{x}_i|_+, |\mathbb{E}[x]-\tilde{x}_i|_-$, and $\gamma_i \tilde{x}_i\left( \sum_{j\in\mathcal{I}} w_{ij}\tilde{x}_j - \tilde{x}_i \right)$.

\subsection{Simulation}
We let $c_i=2$, $n=10$ as in Section \ref{sec:comparison_pricing_schemes} and analyze all three network topologies depicted in Figure \ref{fig:basic_networks}. $a_i$ and $b_i$ are sampled from the same uniform distributions. We set the exogenous price as the profit-maximizing price in the absence of peer effects \eqref{eq:optimal_uniform_price}, from which the expected consumption $\mathbb{E}[x]$ is determined with \eqref{eq:optimal_expected_consumption_unif_price}. The analytical solution to the MIQCP \eqref{eq:utility_targeting_optimization_problem} is determined with \texttt{Gurobi} \cite{Gurobi-Optimization:2016aa}. We repeat this calculation 10,000 times and take the mean across all iterations. To describe the performance of the heuristic, we define the performance metric $S$ as follows:
\begin{align}\label{eq:suboptimality}
S_m = \frac{\Pi_m^h - \Pi^E}{\Pi_m^\ast - \Pi^E}\cdot 100\%,
\end{align}
where $\Pi_m^\ast$ and $\Pi_m^h$ denote the profit under the analytical solution of \eqref{eq:utility_targeting_optimization_problem} and the heuristic with $m$ targeted users, respectively. $\Pi^E$ denotes the profit in the absence of any peer effects ($m=0$) achieved with exogenous price $p$ where the users consume according to $\tilde{\mathbf{x}} = B^{-1}(\mathbf{a}-p\mathbf{1})$. $S_m$ captures the fraction of the heuristic's achieved profit improvement of the total possible improvement.

\begin{figure}[hbtp]
\centering
\vspace*{-0.2cm}
\includegraphics[scale=0.345]{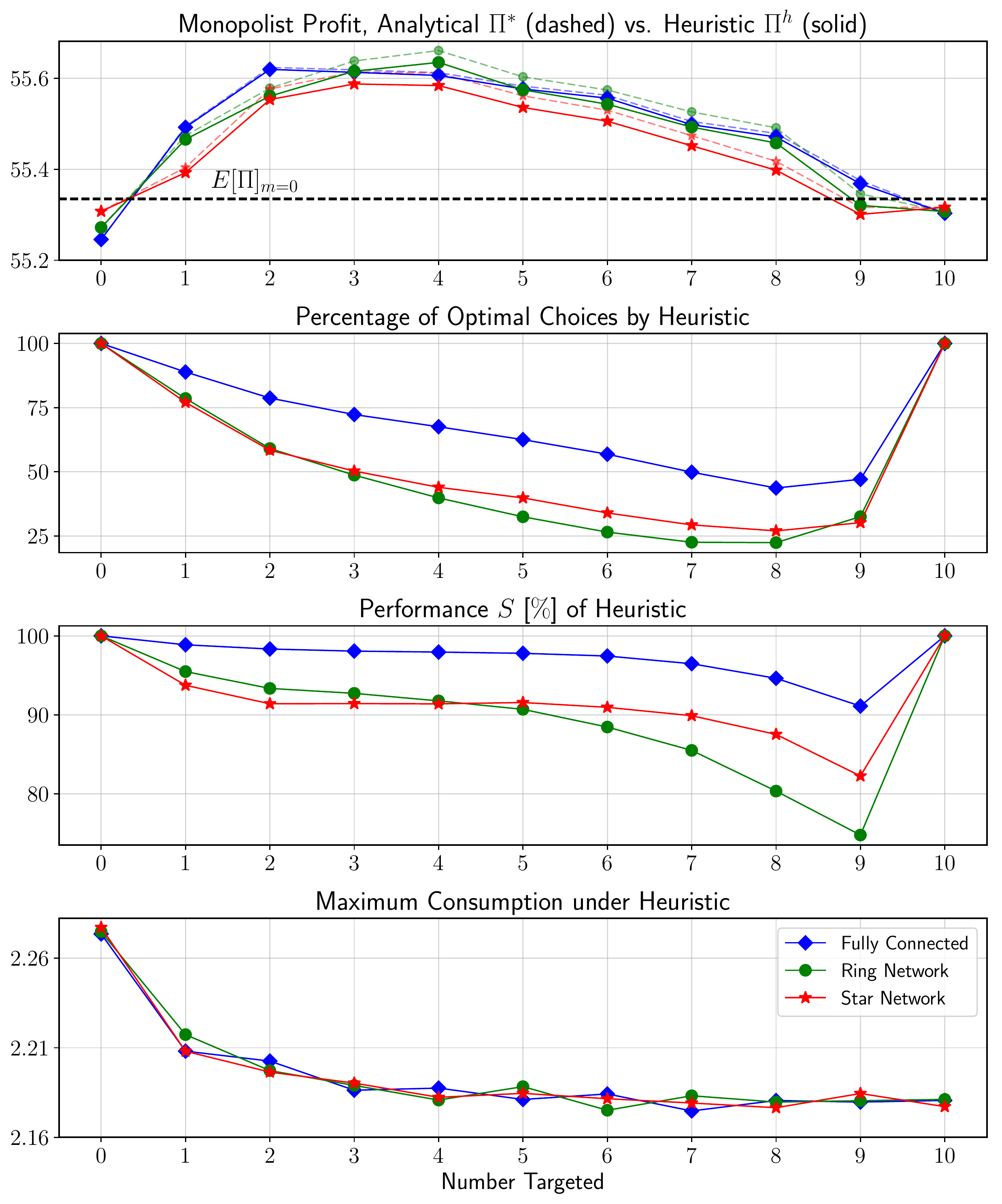}
\caption{Average profit, percentage of optimal choice of heuristic, regret, and average infinity norm of consumption for the utility's profit maximization problem under the single price \eqref{eq:unif_price_complete}. 10,000 iterations, $a\sim\text{unif}[8,12],~b\sim\text{unif}[0.75, 1.25],~c_i=2$.}
\vspace*{-0.1cm}
\label{fig:Profit_Maximization_FixedPrice}
\end{figure}

Figure \ref{fig:Profit_Maximization_FixedPrice} shows the objective for the heuristic $\Pi^h$ (solid lines) and analytical solution $\Pi^\ast$ (colored dashed line) for all network topologies as a function of $m$. The expected profit with $m=0$ follows by taking the expectation of the profit
\begin{align*}
\mathbb{E}[\Pi]_{m=0} =n\cdot\mathbb{E}_{a\sim U[8,12]}\mathbb{E}_{b\sim U[0.75, 1.25]}\left[px - cx^2\right]\Big|_{x=\frac{a-p}{2b}},
\end{align*}
which is depicted as the black dashed line. Further, the percentage of cases where the heuristic selects the identical subset of users as the analytical solution is depicted in the second subplot. $S_m$ and the maximum user consumption as a function of $m$ are provided in the third and fourth subplot, respectively.

For all network topologies, it can be seen that the optimal solution to \eqref{eq:utility_targeting_optimization_problem} achieves an increase in profit by $\approx 1\%$ for $m\in\lbrace 2, 3, 4\rbrace$ compared to the case of no targeting, while at the same time reducing the peak consumption by $\approx 4\%$. The performance of the heuristic decreases in the number of consumers targeted and reaches its minimum at $\approx 75\%$, $\approx 82\%$, and $\approx 90\%$ for the ring, star, and fully connected network, respectively. The percentage of optimal choices across all 10,000 iterations is always $> 22\%$. These results suggest that the presented heuristic achieves a good approximation of the optimal solution, which is NP-hard and computationally intractable for larger, real-world networks.



%
%


\section{Conclusion}
\label{sec:Conclusion}
Motivated by home energy reports that benchmark the consumption of individual users against their neighbors, we proposed a two-stage game-theoretic model for a network of electricity consumers, in which each consumer seeks to optimize her individual utility function that includes a peer effect term. Specifically, users derive positive utility from consuming less energy than the average of their neighbors, and vice versa. We investigated profit-maximizing pricing schemes for the complete and incomplete information scenario as well as for the single price and perfect price discrimination case. We provided theoretical statements with regard to overall consumption, efficiency, and profit under network uncertainty. For the case of targeting only a subset of all available consumers under an exogenous single price, we formulated the monopolist's profit maximization problem. The resulting NP-hard optimization problem was solved with a heuristic approach, which simply targets those users who deviate most from the expected consumption in the hypothetical absence of peer effects. Compared to the analytical solution, this heuristic was shown to achieve acceptable accuracy.

This work could be extended by incorporating time. In particular, if we allow the monopolist to also procure electricity from the wholesale market whose prices are fluctuating, an algorithmic and online treatment of this problem becomes necessary. The goal then becomes to learn user preferences and the network structure over time. Further, the selection problem to target the most valuable users for the objective of profit maximization calls for modeling peer effects in auction settings, where the desired goal is to design a truthful and incentive compatible mechanism to elicit user preferences.


\bibliographystyle{IEEEtran}
\bibliography{bibliography}

%
%


\section*{Appendix}
\label{sec:Appendix}
\subsection*{Proof of Theorem \ref{thm:second_stage_equilibrium}}
With Assumption \ref{as:a_strictly_greater_than_p}, \eqref{eq:consumption_equilibrium} follows by evaluating the first order optimality condition of \eqref{eq:user_utility_function_peer_effect} and acknowledging that its second derivative is strictly negative. Uniqueness of the Nash Equilibrium follows from Topkis' Theorem on supermodular games \cite{Topkis:1998aa}, which holds due to the continuity of the payoff functions \eqref{eq:user_utility_function_peer_effect} on the compact set $\mathbb{R}_+$ and increasing differences in $\left(x_i, \mathbf{x}_{-i}\right)$ as $\frac{\partial^2 u_i}{\partial x_i \partial x_j} \geq 0~\forall~i,j\in\mathcal{I}$.

\subsection*{Proof of Theorem \ref{thm:opt_price_under_perfect_price_discrimination}}
\eqref{eq:optimal_price_under_ppd} is obtained by solving
\begin{equation}\label{eq:optimization_problem_utility}
\begin{aligned}
& \underset{\mathbf{p}}{\text{maximize}}
& & \mathbf{p}^\top \mathbf{x} - \mathbf{x}^\top C\mathbf{x}\\
& \text{subject to}
& & \mathbf{x} = (B + 2\Gamma -\Gamma W)^{-1}\left(\mathbf{a}-\mathbf{p}\right)
\end{aligned}
\end{equation}
and applying the Matrix Inversion Lemma for general matrices $A, U, C, V$ of appropriate dimensions:
\begin{equation*}
(A + UCV)^{-1} = A^{-1} - A^{-1} U\left( C^{-1} + VA^{-1} U \right)^{-1} VA^{-1}.
\end{equation*}
The optimal profit $\Pi^\ast$ is obtained by plugging $\mathbf{p}^\ast$ and $\mathbf{x}^\ast$ into the utility function of the monopolist.

\subsection*{Proof of Theorem \ref{thm:profit_maximizing_uniform_price}}
\eqref{eq:unif_price_complete} is obtained in the same fashion as \eqref{eq:optimal_price_under_ppd}:
\begin{equation*}
\begin{aligned}
& \underset{p}{\text{maximize}}
& & p\mathbf{1}^\top \mathbf{x} - \mathbf{x}^\top C\mathbf{x}\\
& \text{subject to}
& & \mathbf{x} = (B + 2\Gamma -\Gamma W)^{-1}\left(\mathbf{a}-p\mathbf{1}\right)
\end{aligned}
\end{equation*}
Eliminating $\mathbf{x}$ from both equations and evaluating the first order optimality condition with respect to $p$ yields \eqref{eq:unif_price_complete}.

\subsection*{Proof of Theorem \ref{thm:opt_price_incomplete_information}}
To derive \eqref{eq:unif_price_incomplete}, we first note that since $W=W^\top$ and $C=cI$, the profit maximizing solution under complete information \eqref{eq:optimal_price_under_ppd} simplifies to
\begin{align*}
p_u^\ast = \frac{\mathbf{1}^\top\mathbf{a}}{2n} + \mathbf{1}^\top \left[ 2\Gamma + B + cI - \Gamma W \right]^{-1} \frac{\mathbf{a}c}{2n}.
\end{align*}
After taking the expectation with respect to the random variables $\lbrace a_i\rbrace_{i=1}^n$ and $\lbrace b_i \rbrace_{i=1}^n$ to obtain
\begin{equation*}
\tilde{p}_u^\ast = \frac{\mathbb{E}[a]}{2} + \frac{\mathbb{E}[a]c}{2n}\mathbb{E}_b\left[ \mathbf{1}^\top \left[ 2\Gamma + B + cI - \Gamma W \right]^{-1}\mathbf{1}^\top  \right],
\end{equation*}
we first show convexity of the last term in $\text{diag}(2b_1, \ldots, 2b_n)$. Define the matrices
\begin{align*}
D &= \Gamma + \frac{c}{2}I - \frac{\Gamma W}{2} + \alpha~\text{diag}\left(2b_1, \ldots, 2b_n\right),\\
E &= \Gamma + \frac{c}{2}I - \frac{\Gamma W}{2} + (1-\alpha)~\text{diag}\left(2\bar{b}_1, \ldots, 2\bar{b}_n\right),
\end{align*}
where $\alpha \in (0,1)$. $D$ and $E$ are clearly positive definite due to the Levy-Desplanques Theorem \cite{Horn:2013aa}. It is then to be shown that 
\begin{align*}
g(X) &:= \mathbf{1}^\top X^{-1} \mathbf{1},~ X := \left( \alpha D + (1-\alpha)E \right)^{-1},\\
X &:= 2\Gamma + B + cI - \Gamma W
\end{align*}
is a convex function on the domain of all positive definite matrices. Using the Schur Decomposition, which states
\begin{align*}
\begin{bmatrix}
S & T^\top \\ T & U
\end{bmatrix}\succeq 0 \Leftrightarrow S \succeq T^\top U^{-1}T,
\end{align*}
and since positive definite matrices are convex,
\begin{align*}
&\quad\alpha \begin{bmatrix}
\mathbf{1}^\top D^{-1} \mathbf{1} & \mathbf{1}^\top \\ \mathbf{1} & D
\end{bmatrix} + 
(1-\alpha)\begin{bmatrix}
\mathbf{1}^\top E^{-1} \mathbf{1} & \mathbf{1}^\top \\ \mathbf{1} & E
\end{bmatrix} \\
&= \begin{bmatrix}
\alpha\mathbf{1}^\top D^{-1} \mathbf{1} + (1-\alpha) \mathbf{1}^\top E^{-1} \mathbf{1} & \mathbf{1}^\top\\ \mathbf{1} & \alpha D + (1-\alpha)E
\end{bmatrix} \succeq 0,
\end{align*}
This immediately shows convexity of $g(X)$:
\begin{align*}
\alpha g(D) + (1-\alpha)g(E) = \alpha \mathbf{1}^\top D^{-1} \mathbf{1} + (1-\alpha)\mathbf{1}^\top E^{-1}\mathbf{1}\\
\geq \mathbf{1}^\top (\alpha D + (1-\alpha)E)^{-1} \mathbf{1} = g(\alpha D + (1-\alpha)E).
\end{align*}
Finally, applying Jensen's inequality in the multivariate case on the multivariate random variable $Y:=\text{diag}(2b_1,\ldots, 2b_n)$, we obtain
\begin{align*}
\mathbb{E}_Y\left[ g(X) \right] \geq g\left(\mathbb{E}_Y[X]\right),
\end{align*}
from which \eqref{eq:unif_price_incomplete} follows directly.

\subsection*{Proof of Theorem \ref{thm:opt_price_uniform_price}}
Under the given conditions, it follows immediately that $\mathbb{E}[x_1^\ast] = \ldots = \mathbb{E}[x_n^\ast].$ With this constraint, taking the expectation of \eqref{eq:consumption_equilibrium} yields $\mathbb{E}[x^\ast(p)]$ as a function of $p$. Plugging $\mathbb{E}[x^\ast(p)]$ into the utility's profit function \eqref{eq:utility_profit_function} and taking the expectation with respect to $a$ and $b$ allows to compute the optimal uniform price $p^\ast$ \eqref{eq:optimal_expected_consumption_unif_price}. Next, setting $p=p^\ast$ in $\mathbb{E}[x^\ast(p)]$ yields \eqref{eq:optimal_expected_consumption_unif_price}.

\subsection*{Proof of Theorem \ref{thm:optimal_consumption_decreasing_gamma}}
Taking the derivative of \eqref{eq:consumption_equilibrium_rewritten} with respect to $\gamma$ yields:
\begin{equation*}
\frac{d\mathbf{x}}{d\gamma} = -\frac{1}{4\gamma(b+\gamma)}K^{-1} F^{-1}(\mathbf{a}-\mathbf{p}),\quad \gamma>0
\end{equation*}
where we used the abbreviations
\begin{align*}
K :=\left(I - \frac{\gamma}{(2b + 2\gamma)} W \right),~ F :=\left(I + \frac{b}{\gamma}(2I-W)^{-1} \right).
\end{align*}
$K$ is a strictly diagonally dominant \textit{M-Matrix} because it can be expressed in the form $sI-B$ with $s=1$ and has negative off-diagonal elements \cite{Berman:1994aa}. This special property guarantees that its inverse exists and is strictly diagonally dominant and entrywise positive. $F$ is strictly diagonally dominant with positive off-diagonal entries, because $(2I-W)^{-1}$ is an \textit{M-Matrix}. The Levy-Desplanques Theorem \cite{Horn:2013aa} then implies that $F^{-1}$ exists, is diagonally dominant, and possesses nonnegative diagonal elements. Despite the possible negativity of its off-diagonal elements, we show that the row sums of $K^{-1}F^{-1}$ are positive. Take, for example the $i$-th row sum:
\begin{align*}
\sum_{j=1}^n(K^{-1}F^{-1})_{ij} = \sum_{j=1}^n \sum_{s=1}^n K_{is}^{-1}F_{sj}^{-1} =\sum_{s=1}^n K_{is}^{-1}F_{ss}^{-1} + \\
\sum_{s=1}^n K_{is}^{-1}\sum_{j=1,j\neq s}^n F_{sj}^{-1} > \sum_{s=1}^n K_{is}^{-1}F_{ss}^{-1} - \sum_{s=1}^n K_{is}^{-1}~F_{ss}^{-1} = 0.
\end{align*}
Together with $a_i>p_i~\forall~i\in\mathcal{I}$ (see Assumption \ref{as:a_strictly_greater_than_p}), this shows that $\frac{d\mathbf{x}}{d\gamma}<0$ for $\gamma > 0$.

\subsection*{Proof of Theorem \ref{thm:high_consumer_influence}}
Define $L:=(2I-W)$, which is a diagonally dominant matrix.
Evaluating $\frac{d\mathbf{x}}{d\gamma}$ at $\gamma=0$ yields
\begin{align}\label{eq:dx_dgamma_at_zero}
\frac{d\mathbf{x}}{d\gamma}\Big|_{\gamma=0} = -\frac{1}{4}(2I-W)(\mathbf{a}-\mathbf{p}) = -\frac{1}{4}L\boldsymbol{\alpha},
\end{align}
where $\boldsymbol{\alpha}$ is the column vector of all $\lbrace\alpha_i~|~i\in\mathcal{I}\rbrace$. Evaluating this derivative for user $i\neq j,~i\in\mathcal{C}_j$ yields
\begin{align*}
-4\frac{dx_i}{d\gamma}\Big|_{\gamma=0} &= L_{ii}\alpha + L_{ij}\bar{\alpha} + \sum_{k\in\mathcal{I}\setminus\lbrace i,j \rbrace}L_{ik}\alpha \\
&= 2\alpha-\frac{\bar{\alpha}}{n-1} - \frac{(n-2)\alpha}{n-1} \\
&< 2\alpha-\frac{n\alpha}{n-1} - \frac{(n-2)\alpha}{n-1} = 0.
\end{align*}
Hence we have $\frac{dx_i}{d\gamma}\Big|_{\gamma=0} > 0$. On the other hand, for the ``high'' consumer $j$, the derivative reads
\begin{align*}
-4\frac{dx_j}{d\gamma}\Big|_{\gamma=0} &= L_{jj}\alpha + \sum_{k\in\mathcal{I}\setminus j}L_{jk}\alpha = 2\alpha - \frac{n-2}{n-1}\alpha > 0,
\end{align*}
which completes the proof.

\subsection*{Proof of Lemma \ref{lem:restate_high_consumer_influence}}
The proof is similar to the one used for Theorem \ref{thm:high_consumer_influence}. For each user $i$, $i\in\mathcal{I}\setminus j$, the derivative reads
\begin{align*}
-4\frac{dx_i}{d\gamma}\Big|_{\gamma=0} &= 2\alpha - \frac{k\alpha}{m_i} - \frac{m_i-2}{m_i-1}\alpha = \frac{\alpha(m_i-k)}{m_i-1}\\
&\leq \frac{(k-1)-k}{m_i-1}\alpha < 0.
\end{align*}
For user $j$, we have
\begin{align*}
-4\frac{dx_j}{d\gamma}\Big|_{\gamma=0} &= 2\alpha - \frac{\alpha}{m_j-1} = \frac{2m_j-1}{m_j-1}\alpha > 0.
\end{align*}

\subsection*{Proof of Theorem \ref{thm:targeted_peer_effects}}
From Theorem \ref{thm:opt_price_uniform_price}, any user $j$ with index $3, \ldots, n$, given the price $p$, consumes $(a_j - p) / (2b_j)$. To find $x_1^\ast$ and $x_2^\ast$, we solve \eqref{eq:consumption_equilibrium} for users 1 and 2:
\begin{align*}
\begin{bmatrix}
2(b_1 + \gamma) & -\gamma w_{12} \\ -\gamma w_{21} & 2(b_2 + \gamma)
\end{bmatrix}
\begin{bmatrix}
x_1^\ast \\ x_2^\ast
\end{bmatrix}
=
\begin{bmatrix}
a_1 - p + \gamma\sum_{j=3}^n w_{1j}x_j\\
a_2 - p + \gamma\sum_{j=3}^n w_{2j}x_j
\end{bmatrix}
\end{align*}
Comparing $x_1^\ast + x_2^\ast$ to the consumptions without peer effect, that is, $(a_1 - p) / (2b_1) + (a_2 - p) / (2b_2)$ yields the desired inequalities. For the special case $n=2$, note that $w_{12} = w_{21} = 1$ and $w_{2j},~j\geq 3$ as well as $w_{1j},~j\geq 3$, are zero.

\subsection*{Proof of Theorem \ref{thm:uncertainty_network_structure}}
The optimal pricing vector $\mathbf{\tilde{p}}^\ast$ under network uncertainty and its corresponding consumption vector $\mathbf{\tilde{x}}^\ast$ can be determined by solving \eqref{eq:optimization_problem_utility} (with $W=\tilde{W}$) with respect to $\mathbf{p}$. $\mathbf{\tilde{x}}^\ast$ is then determined by plugging $\mathbf{\tilde{p}}^\ast$ back into \eqref{eq:consumption_equilibrium}. Let $F := B + 2\Gamma - \Gamma W$, $\tilde{F}:=\lambda + 2\Gamma - \Gamma \tilde{W}$. Then $\mathbf{\tilde{p}}$ and $\mathbf{\tilde{x}}$ are
\begin{align*}
\mathbf{\tilde{p}}^\ast &= \mathbf{a} - \tilde{F}(C+\tilde{F})^{-1}\mathbf{a}/2,\\
\mathbf{\tilde{x}}^\ast &= F^{-1}\tilde{F}(C+\tilde{F})^{-1}\mathbf{a}/2.
\end{align*}
The optimal profit $\tilde{\Pi}^\ast = \mathbf{\tilde{p}}^{\ast\top} \mathbf{\tilde{x}}^\ast - \mathbf{\tilde{x}}^{\ast\top} C \mathbf{\tilde{x}}^\ast$ can then be expressed as follows:
\begin{align*}
\tilde{\Pi}^\ast = \frac{1}{4}\mathbf{a}^\top (C+\tilde{F})^{-1}\mathbf{a} + \mathcal{O}(\gamma^2) \geq \frac{1}{4}\mathbf{a}^\top (C+\tilde{F})^{-1}\mathbf{a}.
\end{align*}
Using the definition of Rayleigh quotients \cite{Horn:2013aa}, we thus obtain the following ratio on the profit under uncertainty:
\begin{align*}
\frac{\tilde{\Pi}^\ast}{\Pi^\ast} \geq \frac{\mathbf{a}^\top (C+\tilde{F})^{-1}\mathbf{a}}{\mathbf{a}^\top (C+F)^{-1}\mathbf{a}} \geq \frac{\lambda_{\mathrm{min}}((C+\tilde{F})^{-1})}{\lambda_{\mathrm{max}}((C+F)^{-1})}.
\end{align*}
$(C+\tilde{F})$ as well as $(C+F)$ are symmetric positive definite matrices due to their diagonal dominance with nonpositive off-diagonal elements. Hence the eigenvalues of their inverses are strictly positive. Utilizing the identity $\lambda_{\mathrm{min}}(A)^{-1} = 1/\lambda_{\mathrm{max}}(A)$ for any nonsingular matrix $A$, and $\Vert A+B\Vert \leq \Vert A\Vert+\Vert B\Vert$ (a fundamental property of matrix norms), further simplifications yield
\begin{align*}
\frac{\tilde{\Pi}^\ast}{\Pi^\ast} &\geq \frac{\lambda_{\mathrm{min}}(C+F)}{\lambda_{\mathrm{max}}(C+\tilde{F})} = \frac{\lambda_{\mathrm{min}}(C+F)}{\Vert C + F + \gamma (W-\tilde{W}) \Vert_2}\\
&\geq \frac{\lambda_{\mathrm{min}}(C+F)}{\lambda_{\mathrm{max}}(C+F) + \gamma\Vert (W-\tilde{W}) \Vert_2},
\end{align*}
where we used the fact that for a symmetric positive definite matrix $A$, we have $\Vert A\Vert_2 \equiv \sqrt{\lambda_{\mathrm{max}}(A^\top A)} = \sqrt{\lambda_{\mathrm{max}}(A^2)} = \lambda_{\mathrm{max}}(A)$.

\subsection*{Proof of Theorem \ref{thm:efficiency_consumption_equilibrium}}
The social welfare $\mathcal{S}$ is the sum of all users' and the monopolist's utility:
\begin{align*}
\mathcal{S} = \sum_{i\in\mathcal{I}}\left( a_i x_i - b_i x_i^2 - c_i x_i^2 + \gamma_i x_i \left( \sum_{j\in\mathcal{I}}w_{ij}x_j - x_i \right) \right).
\end{align*}
For each $i\in\mathcal{I}$, minimizing $\mathcal{S}$ with respect to $x_i$ yields
\begin{align*}
\frac{d\mathcal{S}}{dx_i} = a_i - 2(b_i + c_i + \gamma_i)x_i + \gamma_i\sum_{j\in\mathcal{I}} w_{ij} x_j + \gamma_i \sum_{j\in\mathcal{I}} w_{ji} x_i,
\end{align*}
where the last term on the right hand side signifies the externalities user $i$ imposes on its neighbors, but which are unaccounted for in the individual users' utility maximization.
Solving for $x_i$ and vectorizing the equation yields \eqref{eq:socially_optimal_consumption}.

To show that $\mathbf{x}_i^o > \mathbf{x}_i^\ast$ for $\gamma > 0$, it suffices to show that $A=(C + B/2 + \Gamma - W^\top\Gamma/2 - \Gamma W/2)^{-1}$ is entrywise greater that $B=(C + B + 2\Gamma - W^\top\Gamma/2 - \Gamma W/2)^{-1}$.
By performing Gauss-Jordan Elimination on $A$ and $B$ and exploiting the fact that $A$ and $B$ are diagonally dominant matrices with positive values on the diagonal and negative off-diagonal entries, this claim follows.

To show that a Pigouvian Subsidy of $s_i = \frac{1}{2}(b_i + \gamma_i)x_i^2$ restores social welfare, note that the user's utility function $u_i^o$ now reads
\begin{align*}
u_i^o = a_i x_i - \frac{1}{2}b_i x_i^2 - p_i x_i + \gamma_i x_i\left( \sum_{j\in\mathcal{I}}g_{ij}x_j - \frac{1}{2}x_i \right).
\end{align*}
The solution to the subgame-perfect equilibrium under the new user utility $u_i^o$ yields $x_i^o$.


\end{document}